\newcommand{\ergs}{\mbox{erg\,s$^{-1}$}}

\documentclass[referee]{cjaa}           

\usepackage{graphicx}                   
\input{epsf.sty}                        
\input{psfig.sty}

\begin{document}


\title{Bright X-ray source populations in the
starburst galaxies NGC 4038/4039}


\volnopage{Vol.0 (200x) No.0, 000--000}      
   \setcounter{page}{1}          

   \author{Xi-Wei Liu
   \and Xiang-Dong Li
      }
   \offprints{Xi-Wei Liu}                   

   \institute{Department of Astronomy, Nanjing University, Nanjing 210093,
China\\
             \email{liuxw@nju.edu.cn, lixd@nju.edu.cn}
          }

   \date{Received~~; accepted~~}

\abstract{Assuming a naive star formation history, we construct the
synthetic X-ray source populations for comparison with the X-ray
luminosity function (XLF) of the interacting galaxies NGC 4038/4039
using a population synthesis code. We have considered high- and
intermediate-mass X-ray binaries, young rotation-powered pulsars and
fallback disk-fed black holes in modeling the bright X-ray sources
detected. We find that the majority of the X-ray sources are likely
to be intermediate-mass X-ray binaries, but for typical binary
evolution parameters, the predicted XLF seems to be steeper than
observed. We note that the shape of the XLFs depends critically on
the existence of XLF break for young populations, and suggest
super-Eddington accretion luminosities or the existence of
intermediate-mass black holes to account for the high luminosity end
and the slope of the XLF in NGC 4038/4039. 
}

\authorrunning{Liu \& Li }            
   \titlerunning{X-ray source populations in NGC 4038/4039 }  

   \maketitle

\section{Introduction}

With {\it Chandra}'s unprecedent sensitivity and angular resolution,
populations of individual X-ray sources, with luminosities
comparable to those of Galactic X-ray binaries, can be detected at
the distance of the Virgo Cluster and beyond (Fabbiano 2006). {\it
Chandra} observations of the merging galaxies NGC 4038/4039 (the
Antennae) provide a unique opportunity for the study of the X-ray
source populations in young and intense starburst environments
(Fabbiano, Zezas \& Murray \cite{fab01}; Zezas et al.
\cite{zezas02a,zezas02b}; Zezas \& Fabbiano \cite{zezas02},
hereafter ZF02). Among the 43 point-like sources detected by {\it
Chandra} (down to a limiting luminosity of $\sim 10^{38}\,\ergs$),
17 sources have inferred luminosities above $\sim 10^{39}\,\ergs$,
assuming an isotropic emission. These luminosities exceed the
Eddington limit of a $10\,M_{\sun}$ accreting black hole (BH), and
we classified these sources as ultraluminous X-ray sources (ULXs).
Although accretion binaries with a BH of mass in the range
$100-1000M_{\odot}$ (i.e., intermediate mass black holes, IMBHs) can
easily explain these super-Eddington luminosities, they may not
account for the majority of the ULXs in the Antennae galaxies
(ZF02). One possible explanation is that the X-ray radiation of the
ULXs is not isotropic, so the real luminosities are much lower the
inferred values (King et al. 2001). Alternatively, Begelman
(\cite{begelman02}) suggested that in radiation-dominated accretion
disks, the radiation could escape from the disk (due to photon
bubble instability) at a rate higher than predicted by the standard
accretion disk theory, so that the escaping flux could exceed the
Eddington luminosity by a factor of up to $\sim 10-100$. ZF02 and
Zezas et al. (2007) derived the X-ray luminosity function (XLF) of
the X-ray source populations in the Antennae. The cumulative XLF is
well fit by a power-law with an index of $\sim 0.5-0.8$, consistent
with (though slightly flatter than) the slope of the ``universal
luminosity function" for the HMXB populations in various types of
galaxies (Grimm, Gilfanov \& Sunyaev \cite{grimm03}). Here, we model
the observed XLF (with both the shape and the absolute source
number) by using evolutionary population synthesis (EPS) method, to
evaluate the effects of the input parameters on the calculated
results. Although there have been many observations of the point
X-ray sources in external galaxies, theoretical investigations on
the X-ray source populations remain to be limited. Ghosh \& White
(2001) discussed the evolution of X-ray luminosities of normal
galaxies with the cosmological evolution of the star formation rate
(SFR) in a semi-numerical fashion. Wu (2001) constructed an analytic
model and calculated the luminosity function of X-ray binaries in
external galaxies. Both of them have adopted simple assumptions on
the formation and evolution of X-ray binaries. A recent population
synthesis study on the XLF of NGC 1569 was done by Belczynski et al.
(\cite{belczynski04}).

This paper is organized as follows. In \S 2 we describe the
population synthesis method and the models for various types of
X-ray sources. The calculated results are presented in \S 3.
Discussion and conclusions are in \S 4.

\section{Model description}

\subsection{Method and input parameters}

We use the EPS code developed by Hurley, Pols, \& Tout
(\cite{hurley00}) and Hurley, Tout, \& Pols (\cite{hurley02}) to
calculate the expected numbers and X-ray luminosity distributions in
the Antennae for various types of single and binary X-ray source
populations.  This code incorporates the evolution of single stars
with binary star interactions, such as mass transfer, mass
accretion, common-envelope (CE) evolution, collisions, supernova
(SN) kicks, tidal friction and angular momentum loss mechanisms
(i.e. mass loss, magnetic braking and gravitational radiation).  In
this work, we have modified the EPS code so that it can perform
better in simulating binary evolution in a variety of circumstances
(see appendix).

We start with the parameters input into the code for population
synthesis. We assume a fraction $f$ ($0<f<1$) of stars are born in
binaries. Surveys of M dwarfs within 20 pc from the Sun have
suggested that $f$ may be a function of stellar spectral types
(Fischer \& Marcy \cite{fischer92}). According to recent works on
stellar multiplicity (Lada \cite{lada06}; Kobulnicky, Fryer \&
Kiminki \cite{kobulnicky06}), stars with later spectral types are
more likely single, for example, $f>50\%$ for G stars, and $f>0.6$
(in the most probable range $0.7<f<0.85$) for massive O/B stars in
the Cygnus OB2 association surveyed by KFK06. We have adopted
$f=0.5$ or 0.8 in our calculations. The initial mass function (IMF)
of Kroupa, Tout, \& Gilmore (\cite{Kroupa}, hereafter KTG93) is
taken for the primary's mass ($M_1$) distribution.  For the
secondary stars (of mass $M_2$), in our standard model we assume
uniform distributions of the mass ratio $q=M_2/M_1$ between 0 and 1
and of the logarithm of the orbital separation $(\ln a)$. However,
KFK06 suggested more general power-law distributions of the
secondary masses and orbital separations, i.e., $P(q)\propto
q^{\alpha}$ and $P(\ln a)\propto(\ln a)^{\beta}$, where the
power-law indices $\alpha>0$ and $-2<\beta<0.5$. We have designed
several models to examine the effects of changing these primordial
binary parameters ($f$, $\alpha$ and $\beta$) on the results.

Reliable inputs for the SFR and star formation history (SFH) are
crucial to population synthesis of the X-ray sources detected in the
Antennae galaxies. For the SFR we adopt the value of
$7.1\,M_{\odot}$yr$^{-1}$ for stars more massive than $5\,M_{\odot}$
(Grimm, Gilfanov \& Sunyaev \cite{grimm03}), which was obtained
based on the measurement of several canonical indicators of SFR,
namely UV, H$\alpha$, FIR and thermal radio emission fluxes.  This
value was then converted into the specific SFRs $S_{s}$ and $S_{b}$
used in the EPS code. Here $S_{s}$ and $S_{b}$ (in unit of
yr$^{-1}$) denote the formation rates of single and binary star
populations, respectively.  Since the binary fraction is $f$, we
have
\begin{equation}
   \frac{2S_{b}}{S_{s}+2S_{b}}=f.
\end{equation}
The amount of mass produced per year in stars more massive than
$5\,M_{\odot}$ is
\begin{equation}
S_{b}\int^\infty_5 m_{1}\xi(m_{1})\,dm_{1}+S_{b}\int^\infty_5
\xi(m_{1})\left(\int^{m_{1}}_5
\frac{m_{2}}{m_{1}}\,dm_{2}\right)\,dm_{1}+S_{s}\int^\infty_5
m\xi(m)\,dm=7.1,
\end{equation}
where $\xi(m)$ is the IMF.  The fist and second terms on the
left-hand-side of Eq.~(2) represent the amount of mass formed per
year as the primary and secondary star of a binary, respectively.
Contribution from single stars is evaluated in the third term.
Combining Eqs.~(1) and (2), we have
\begin{equation}
S_{s}=\frac{14.2-14.2f}{0.144-0.045f} \mbox{\ and\  }
S_{b}=\frac{7.1f}{0.144-0.045f}.
\end{equation}
The values of the single and binary star SFRs can be input into the
single stellar evolution (SSE) and binary stellar evolution (BSE)
code, respectively (see \S 4.1 of HTP02 for details). For a binary
fraction $f=0.5$, we get $S_{s}=58.4$\,yr$^{-1}$ and
$S_{b}=29.2$\,yr$^{-1}$.

According to current models on the evolution of the Antennae,
closest passage of the two galaxies comprising the Antennae happened
roughly 200\,Myrs ago (Barnes \cite{barnes88}; Mihos, Bothun \&
Richstone \cite{mihos93}).  The merging induced star formation tends
to persist for 0.5 to 1 dynamical time (i.e., the orbital time-scale
of the last encounter) (Mengel et al. \cite{mengel05}). It is
expected that most of the X-ray sources detected are relatively
young objects produced after the merging process, so we adopt a
naive model for the SFH that assumes the Antennae have been keeping
the current SFR for the last 300 or 100 \,Myr. Since we are
concerned with young populations, we do not take the SFH earlier
into account.

\subsection{Models for X-ray sources}

We consider three types of X-ray source populations  for the X-ray
sources detected in the Antennae.

A large fraction of the X-ray sources should be X-ray binaries
containing accreting neutron stars (NSs) or BHs. For starburst
galaxies like the Antennae, high-mass X-ray binaries (HMXBs) with
the secondary masses $M_{2}>8\,M_{\odot}$ and intermediate-mass
X-ray binaries (IMXBs) with $2\,M_{\odot}<M_{2}<$8\,M$_{\odot}$ are
the most natural explanation for the bright point-like sources in
these galaxies (low-mass X-ray binaries are too old for our assumed
SFH). According to the manner of the accretion onto the accretor
(NSs or BHs), X-ray binaries can be divided into two classes, i.e.,
wind-accreting (WA) and Roche-lobe overflow (RLOF) systems. In the
former case we adopt the standard Bondi \& Hoyle (1944)
wind-accretion formula to calculate the mass accretion rate of the
NSs/BHs. The velocity $v_{\rm W}$ of the winds from the massive
donor stars is taken to be $v_{\rm W}=\sqrt{2\beta (GM_{2}/R_2)}$,
where $R_2$ is the radius of the donor star, and the value of
$\beta$ depends on the spectral type of the stars. HTP02 suggested
it is in the range of $0.125-7.0$. The gravitational energy released
by the accreted material is converted into radiation, and we have to
correct the calculated bolometric luminosities into those in the
{\it Chandra} detection band of $0.1-10$\,keV.  Since the WA NS
systems have relatively hard spectra that can be described with a
$\Gamma\sim0$ power law below 10 keV (e.g. Campana et al.
\cite{Campana}), we assume that almost all the radiation energy is
concentrated in the {\it Chandra} band, and no bolometric correction
is needed.  For the other RLOF X-ray binaries we adopt a correction
factor of 0.5.

Current stellar evolution models predict that during the core
collapse of massive stars, a considerable amount of the stellar
material will fall back onto the compact, collapsed remnants (NSs or
BHs), usually in the form of an accretion disk.  Li (\cite{li03})
suggested that some of the ULXs in nearby galaxies, which are
associated with supernova remnants, may be BHs accreting from their
fallback disks.  Not all natal BHs can possess a fallback disk,
unless the BH progenitors have enough angular momentum to form a
centrifugally supported disk. The condition that matter does not
spiral into the BH at the onset of collapse is that the rotational
energy of the matter should be greater than half the gravitational
potential energy (Izzard, Ramirez-Ruiz \& Tout \cite{izzard04}):
$J^{2}/(2I)\geq GM_{\rm BH}m/(2r_{\rm LSO})$, where $J$ is the
angular momentum of the matter, $I=mr_{\rm LSO}^{2}$ is its moment
of inertia and $r_{\rm LSO}$ is the radius of the last stable orbit
(LSO) around a BH. In fact for Population I ($Z=0.02$) stars, from
our calculations only those in close binaries can be spun up so fast
(due to spin-orbit tidal evolution) that fallback disks can form
after core collapse. After an initial transient phase of duration
$t_0$, the disk mass $M_{\rm d}$, outer radius $R_{\rm d}$, and mass
accretion rate $\dot{M}_{\rm d}$ obey the simple power-law
evolution, i.e. $M_{\rm d}\propto t^{-p}$, $R_{\rm d}\propto t^{2p}$
and $\dot{M}_{\rm d}\propto t^{-(1+p)}$ (Cannizzo, Lee \& Goodman
\cite{cannizzo90}), and we adopt $p=3/16$ in this paper. Menou,
Perna \& Hernquist (\cite{menou01}) argued that the power-law
evolution breaks down when the disk is cool enough to become
neutral.  We thus limit our calculation on the disk evolution to the
neutralization time $t_{\rm n}$. Following MPH01 we derived the time
$t_{\rm n}$ to be
\begin{equation}
t_{\rm n}\simeq 6.6\times10^{\frac{12-35p}{1+7p}}
R_{8}(t_{0})^{\frac{7p-6}{2+14p}}\left[\frac{M_{\rm d}(t_{0})}
{10^{-3}M_{\odot}}\right]^{\frac{1}{1+7p}}T_{\rm
c,6}^{-\frac{7p}{1+7p}} \mbox{ yr,}
\end{equation}
where $R_{8}(t_{0})$ is the initial radius of the disk in units of
10$^{8}$\,cm, and $T_{\rm c,6}$ the typical temperature of the disk
during the initial transient phase in units of 10$^{6}$\,K. Here we
adopt $R_{8}(t_{0})=1$, and $T_{c,6}=1$.  For $M_{\rm d}(t_{0})$ we
assume a uniform distribution of $\log M_{\rm d}(t_{0})$ in the
range of $10^{-5}-1\,M_{\odot}$. The bolometric correction factor of
the fallback disk-fed systems is also chosen to be 0.5.

Moreover, the X-ray emission from accreting compact stars (both
single and in binaries) is generally assumed to be isotropic.
However, for disk accreting sources, the geometric effect may affect
the apparent luminosity distributions (Zhang \cite{Zhang}).  For the
RLOF X-ray binaries and the fallback disk-fed BHs we have taken into
account the geometric effect in calculating their X-ray luminosities
(see also Liu \& Li 2006).

Rotation-powered pulsars also shine in X-rays. There appears to be a
strong correlation between the rates of rotational energy loss
$\dot{E}$ and their X-ray luminosities $L_{\rm X}$ (Seward \& Wang
\cite{Seward}; Becker \& Trumper \cite{Becker}; Possenti et al.
\cite{Possenti}). Generally younger pulsars have stronger X-ray
emission. Perna \& Stella (\cite{perna04}) argued that a fraction of
ULXs could be young, Crab-like pulsars, espeically starburst
galaxies should each have several of these sources, and the X-ray
luminosity of a few percent of galaxies is dominated by a single
bright pulsar with $L_{\rm X}\geq 10^{39}$ ergs$^{-1}$, roughly
independent of its SFR. Following Perna \& Stella (\cite{perna04})
and Liu \& Li (2006) we adopt the empirical relation by Possenti et
al. (2002) to estimate the $2-10$ kev X-ray luminosities of young
pulsars.

To our knowledge this work is the first to include all the three
types of X-ray source populations suggested in the literature in
modeling the XLF of external galaxies.

\section{Results}
We have adopted a variety of models with different assumptions for
the input parameters (see Table 1).  We set Model 1 (hereafter M1)
as the control model, while other models are designed to test the
effects of the input parameters. In Fig.~1 we show the calculated
XLF in M1, and the components contributed by HMXBs, IMXBs, fallback
disk sources, and pulsars in the Antennae.  One can see that IMXBs
dominate the total X-ray luminosity range, and young pulsars also
contribute a few of the brightest sources. HMXBs play a minor role
in the total population due to their relatively short lifetime
compared to IMXBs. Fallback disk sources are much fewer than the
above three classes of sources. The break at $\sim 4\times 10^{38}$
ergs$^{-1}$ is caused by the Eddington luminosity of accreting BHs.

A comparison of the calculated and measured XLFs is shown in Fig.~2.
Obviously the model predicts more X-ray sources at luminosities
$L_{\rm x}<5\times 10^{38}$ ergs$^{-1}$, but less when $L_{\rm
x}>5\times 10^{38}$ ergs$^{-1}$. Especially the break in the XLF
makes it considerably steeper than observed. We note that the XLF
breaks for young populations seem to be a general feature predicted
in theoretical works (e.g. Wu \cite{wu01}; Fig.~1 of Belczynski et
al. 2004; Fig.~6 of Rappaport et al. 2005). Consider that the young
X-ray populations have relatively flat XLF (Grimm et al.
\cite{grimm03}), the analysis on the difference in the modeled and
observational XLFs should not only limited to the Antennae, but may
have a more general application.

The shape of the XLF was suggested to be related to the star
formation activity (Wu \cite{wu01}) and to the NS/BH mass spectrum
and the mass transfer rates in binary stars (Grimm et al. 2003). We
will check these points in the following. There are large
uncertainties in estimating the SFH and SFR in the Antennae.
Obviously, shorter SFH leads to smaller number of XRBs, but
decreasing the duration of star formation in the Antennae seems not
affect the XLF considerably. For example, if the SFH is changed from
300 Myr to 100 Myr, the lower end of the XLF is down by a factor of
$\sim 40$, while the higher end changes little. If we take half of
the adopted SFR (model M3), however, the number of the X-ray sources
with $L_{\rm x}<5\times 10^{38}$ ergs$^{-1}$ can be significantly
decreased to be compatible with observations. This is easily
understood, since the formation of H/IMXBs is closely correlated
with the global SFR. Alternatively, the same effect can be reached
if we lower the binary fraction from $50\%$ in M1 to $10\%$ (model
M2). This is also reasonable, since, if most of the X-ray binaries
originate from the star clusters in the Antennae (Zezas et al.
\cite{zezas02b}), those ejected from the clusters usually have
relatively high space velocities, and are more likely to be
disrupted due to SN kick (Belczynski et al. 2006). The third way to
reduce the number of H/IMXBs is to decrease the CE parameter
$\eta_{\rm CE}$. In model M4, $\eta_{\rm CE}$ is set to be 0.1, much
smaller than unity usually adopted, the mass ratio $q$ and orbital
separation $\ln a$ distributions are taken to be $q^1$ and $(\ln
a)^{-1}$, respectively. These factors can significantly decrease the
number of IMXBs - the peculiar $q$ and $\ln a$ distributions cause a
bias toward the formation of narrow HMXBs, while small $\eta_{\rm
CE}$ leads to coalescence of the progenitors of IMXBs. The resulting
XLF does not show any prominent break since the XLF is now dominated
by young pulsars and the fallback disk sources. This feature can be
tested by future multi-wavelength observations.

However, even with the above modifications, the high end of the XLF,
which plays a critical role in determining the slope of the XLF, is
still not accounted for (see Fig.~2). Our calculations show that it
hardly depends on the IMF (Stolte et al. \cite{stolte02}),
distributions of the mass ratio or orbital separation, and the
metallicity.

Figure 1 clearly suggests that the slope of XLF critically depends
on the Eddington luminosities of BHs in IMXBs. In M1 the actual
maximum luminosities of accreting BHs and NSs are taken to be 10
times the corresponding Eddington luminosities. It has been pointed
out that these values can be as high as $\sim 100$ if the accretion
rates are very high in the cases of H/IMXBs (Belgelman 2002). In
models M2 and M4, we further increase the super-Eddington factor
from 10 to 100, and the XLF correspondingly extends to $\la 10^{40}$
ergs$^{-1}$. The modeled XLFs discussed above are demonstrated in
Fig.~2, in comparison with the observed XLF. The solid, dashed,
dot-dashed, and dotted lines represents the results of models M1-M4,
respectively.

\section{Discussions}

We have performed population synthesis calculations to explore the
possibility of reproducing the observed XLF in the Antennae with
current understanding of formation and evolution of compact stars.
According to our calculations, there are two parameters which affect
the XLF most prominently: the binary formation rate and the factor
of super-Eddington accretion rate allowed. The former parameter,
depending on the global SFR and IMF, the binary fraction and
interactions, determines the overall number of the X-ray sources and
the mass spectrum of compact stars. The latter is related to the
accretion behavior, and constrains the location of the break in the
XLF. They jointly design the shape of the XLF. For the Antennae we
conclude that the majority of the luminous X-ray sources are likely
to be IMXBs, but their formation should be strongly influenced by
star formation activity and binary interactions. Zezas et al.
(\cite{zezas02b}) showed that the position of most of the Antennae
sources is near but {\it not} coincident with optical young clusters
(with typical offsets $\sim 100-300$ pc). Assuming that this
displacement is caused by the motion due to supernova kicks, these
authors suggested that most of the X-ray sources may be accreting
NS/BH binaries with donor stars of masses ranging from 2 to
$10\,M_{\odot}$. This inference is consistent with our result that
IMXBs dominate the XLF in the standard model M1 (see Fig.~1).
However, the XLF in the model shows a clear break at a few
$10^{38}\,\ergs$, which is related to the Eddington luminosities of
accreting BHs. This makes the XLF steeper in the ULX luminosity
range ($>10^{39}$\,\ergs) than the observational one, although the
latter is subject to small number statistics and large observational
uncertainties. There is also evidence for the breaks in the XLFs for
other nearby galaxies (e.g. Sarazin et al. 2000; Shirey et al.
2001). However, the break luminosities are still unclear because of
the distance uncertainties. We find that super-Eddington X-ray
luminosities are required to account for the flat XLF extending to
$\sim 10^{40}$ ergs$^{-1}$. Theoretically it has been realized that
rapid mass accretion is more likely to occur in IMXBs than in LMXBs,
where mass transfer proceeds on thermal rather nuclear timescale. It
is also in accord with the strong correlation found between the
number and average luminosity of ULXs in spirals and their host
galaxy¡¯s far-infrared luminosity (Swartz et al. \cite{swar04}).
Alternatively, some of the most luminous X-ray sources in the
Antennae could be IMBHs. Only a few these objects can interpret the
``overabundance" of the high luminosity end of the XLF. The
available observations are not able to present stringent constraints
on the nature of these X-ray sources, and rule out the possibility
of IMXBs in the Antennae.

 Tennant (\cite{tennant01}) firstly identified the two
different types of XLF in the Galactic disk and bulge: the XLF of
the disk follows a power law with an index $\sim 0.5$ and that of
the bulge follows a similar shape but with a steeper slope. Kilgard
et al. (\cite{kilgard02}) showed that the XLF slope is correlated
with the age of the X-ray binary populations, and the XLFs of
starburst galaxies with relatively more luminous sources are flatter
than those of spiral galaxies.  Wu (\cite{wu01}) constructed a
birth-death model and calculated the XLF of X-ray binaries in
external galaxies.  It was found that the location of the break in
the XLF depends on the look-back time of the previous starburst
activity, and no break appears if there is no starburst (as for the
disk sources in M81). In Kilgard et al. (\cite{kilgard02}), all of
the X-ray sources were assumed to be members of a single population
with uniform properties except for luminosity and lifetime. The
actual situation may be much more complicated. For example, IMXBs
generally form later than HMXBs after the starburst, but they can be
much more brighter than wind-fed HMXBs when RLOF occurs. This mixed
formation sequences of various types of X-ray binaries could produce
``ripples" rather a uniformly evolving break in the XLF. This
conclusion is only for the young X-ray populations (with age
$<300$\,Myr). Since we do not consider low-mass X-ray binaries, we
cannot evaluate the long-term ($\sim$\,Gyr) effect of starburst
here.  It is possible that there exists a uniformly break evolution
in the XLF of old stellar populations (e.g. Kong et al.
\cite{kong02}).

Our population synthesis study is obviously subject to many
uncertainties. Little are known about not only the detail SFR, SFH,
and IMF in the Antennae, but also some key processes, for example,
the CE evolution, in the formation and evolution of X-ray binaries.
Estimation of the X-ray luminosities is also influenced by the
detailed accretion modes in X-ray binaries. Finally, It should be
emphasized that the calculated mass transfer rates (and the X-ray
luminosities) are long-term, averaged ones. It is unclear how to
relate these secular mass transfer rates to observable instantaneous
X-ray luminosities. More detailed X-ray observations of a large
sample of galaxies across the full Hubble sequence are required to
determine the nature of the break luminosities in the XLF and the
XLF evolution. These observations, coupled with development of
stellar and binary evolution models, may provide new insights into
the compact object populations in external galaxies.
%




\begin{acknowledgements}

We would like to thank Jarrod R. Hurley for kindly providing us the
SSE and BSE codes and valuable conversations.  We are grateful to an
anonymous referee for a number of insightful remarks, and Yang Chen,
Hai-Lang Dai, Wen-Cong Chen and Zhao-Yu Zuo for useful diskussions.
This work was supported by Natural Science Foundation of China under
grant numbers 10573010 and 10221001.

\end{acknowledgements}

\clearpage

\begin{figure}
 \includegraphics[width=12cm]{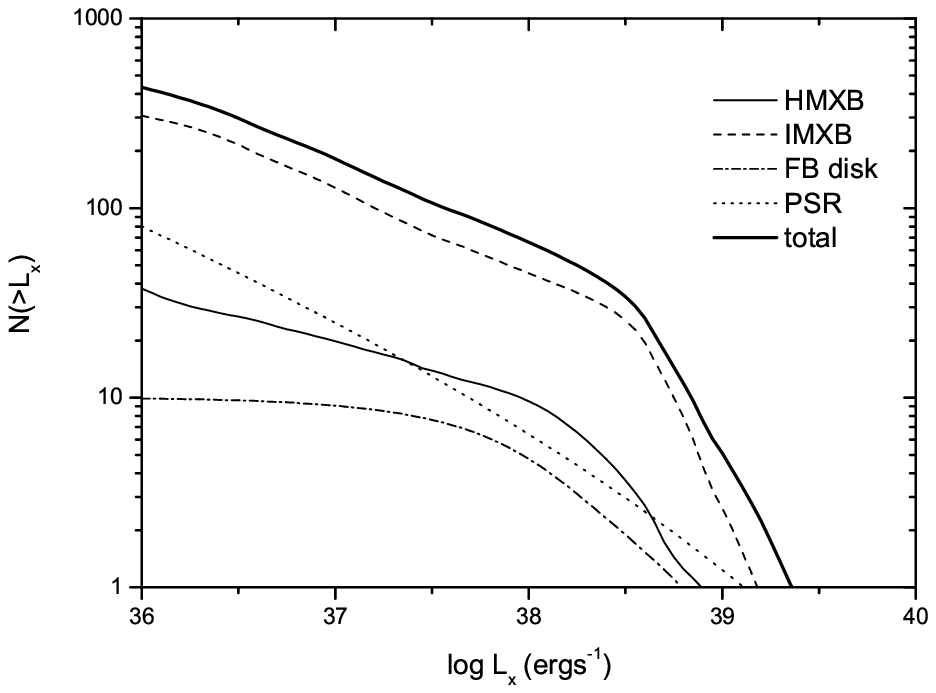}
      \caption{The XLFs and its components in model M1.
              }
\end{figure}

\clearpage

\begin{figure}
 \includegraphics[width=12cm]{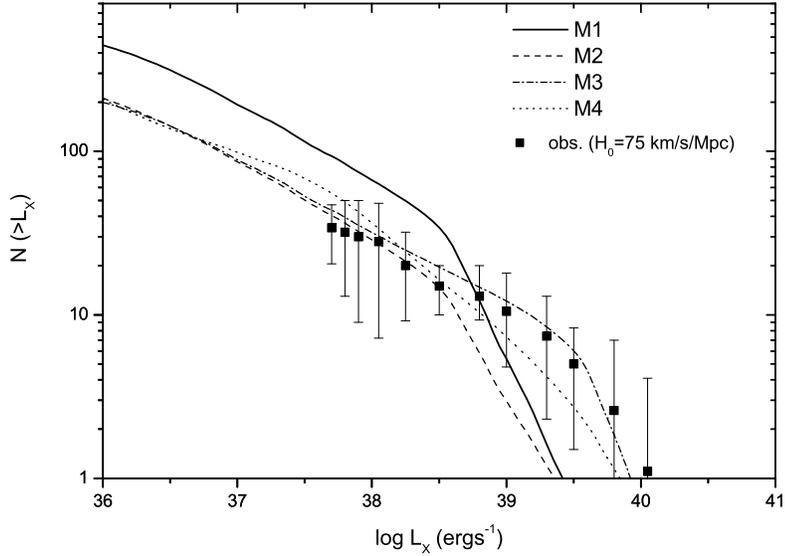}
      \caption{Comparisons of the modeled and observed XLFs. M1 is the standard model.
      Compared with M1, in M2 the binary fraction is decreased by a factor of 5, in M3 the
      SFR is decreased by a factor of 2, and the factor for super-Eddington accretion rate increased
      by a factor of 10 (note that the XLF in M2 can also extend to $\sim 10^{40}\,\ergs$ if the latter
      assumption is adopted). In M4 we take the value of the CE parameter $\eta=0.1$ and
      atypical distributions of mass ratio and orbital separation.}
\end{figure}

%
%
%
%
%

\clearpage

\begin{table*}
\caption{Model Parameters.}
\begin{tabular}{ccccccc}\hline\hline
   Model & $\eta$ & SFR ($M_{\sun}$yr$^{-1}$) & $f$ & $P(q)$ & $P(\ln a)$ & Edd\\ \hline
      M1 & 1.0 & 7.1 & 0.5 & $\propto q^{0}$ & $\propto (\ln a)^{0}$ & 10\\
      M2 & 1.0 & 7.1 & 0.1 & $\propto q^{0}$ & $\propto (\ln a)^{0}$ & 10\\
      M3 & 1.0 & 3.5 & 0.5 & $\propto q^{0}$ & $\propto (\ln a)^{0}$ & 100\\
      M4 & 0.1 & 7.1 & 0.8 & $\propto q^{1}$ & $\propto (\ln a)^{-1}$ & 100\\\hline

\end{tabular}
\smallskip\\
Note: $\eta$ - the CE efficiency parameter; SFR - the SFR for stars
massive than $5\,M_{\sun}$; $f$ - initial binary fraction; $q$ -
initial mass ratio; $a$ - initial orbital separation; Edd - the
factor of super-Edddington accretion rate allowed.

\end{table*}

\clearpage
\appendix                  

\section{Modifications to the original EPS code}

Firstly, the original EPS code produces BHs with unreasonable low
masses. In Fig.~A1 we show the masses of the stellar remnants (i.e.
compact objects) as a function of the masses of their zero-age main
sequence (ZAMS) stars. The BH masses are determined by an empirical
function of the Carbon-Oxygen core masses of the pre-collapse
progenitor stars, i.e. $m_{\rm BH}=1.17+0.09m_{\rm co}$. In the top
panel of Fig.~A1 we show the calculated results with the original
code.  We note that the natal BH masses are in a narrow range of
$\sim1.8-2.0\,M_{\odot}$, which are significantly smaller than those
($\sim 3-18\,M_{\odot}$) of the observed BH candidates in our Galaxy
(Remillard \& McClintock 2006). We have modified the empirical
formula in the code to be $m_{\rm BH}=-33+6m_{\rm co}$, so that a
larger range of BH masses ($\sim 4-10\,M_{\odot}$ and $\sim
5-26\,M_{\odot}$ for population I and II stars respectively) can be
produced, which are shown in the middle and lower panels. The figure
illustrates that the lower limits for the BH progenitor masses are
$\sim 23\,M_{\odot}$ ($Z=0.02$) and $\sim 20\,M_{\odot}$
($Z=0.001$), consistent with the putative values (e.g. Fryer
\cite{fryer99}). It is noted that similar modification was made by
Belczynski, Sadowski \& Rasio (\cite{bsr04}).

Secondly, the EPS code assumes that all non-compact stars suffer
magnetic braking if it is more massive than $0.35\,M_{\odot}$.
However, stars more massive than $1.5\,M_{\odot}$ have radiative
envelopes which cannot generate strong magnetic field.  So we
taken into account the magnetic braking mechanism only for stars
of $M<1.5\,M_{\odot}$. Thirdly, when a main sequence star
overfills its Roche lobe and is 2.5 times more massive than its
companion (the accretor), a very rapid, dynamically unstable mass
transfer phase would emerge, and the orbital separation would
decline dramatically. We thus assume the binary would coalescence,
and no further calculation is taken.

The fourth modification is about the natal BH kick velocities.
During the SN explosions, a kick velocity $v_{\rm k}$ is imparted
on the newborn compact stars with the Maxwellian distribution
\begin{equation}
   P(v_{\rm k})=\sqrt{\frac{2}{\pi}}\frac{v^{2}_{\rm k}}{\sigma^{3}}
   \exp(-\frac{v^{2}_{\rm k}}{2\sigma^{2}})
\end{equation}
where $\sigma=265\,{\rm kms}^{-1}$ for neutron stars (NSs) (Hobbs et
al. \cite{Hobbs}). We assume that only those BHs experienced SN
explosions would be imparted on a natal kick velocity, which is
inversely proportional to the BH mass, i.e. $v_{\rm kick,
BH}=1.4/M_{\rm BH}\times v_{\rm kick, NS}$. Apart from core collapse
SNe, Belczynski \& Taam (\cite{bt04}) suggested the formation of NSs
via accretion-induced collapse of massive white dwarfs. We do not
take it into account due to the large uncertainties of the AIC
assumption.

\clearpage

\begin{figure}
    \begin{center}
    \includegraphics[width=12cm]{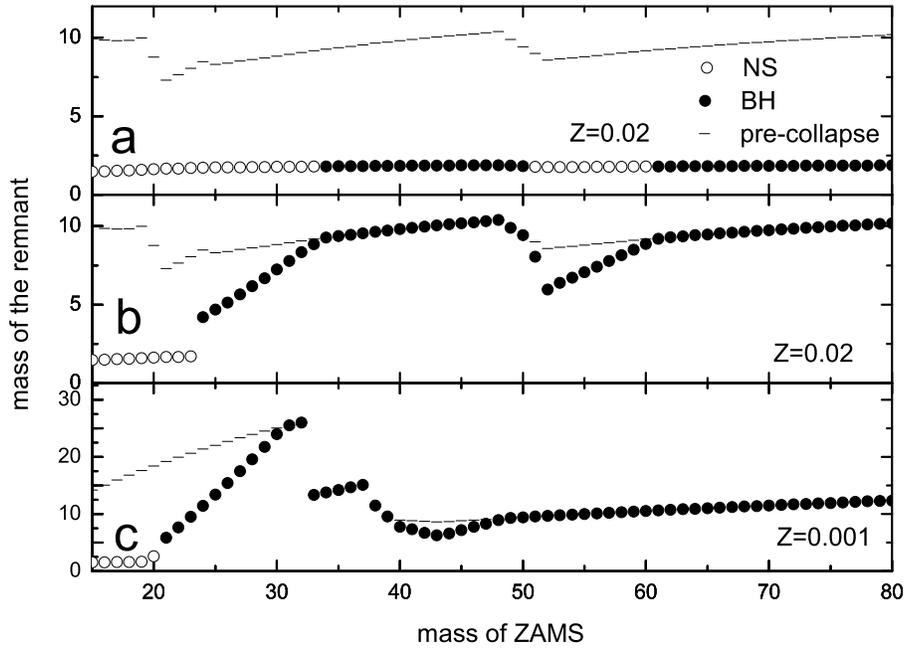}
      \caption{Final remnant (NS/BH) masses as a function of
      ZAMS progenitor mass for different metallicities.  The top panel shows
      the calculated result with the original BSE code, the middle and bottom
      panels show those with our modified code.   }
   \end{center}
\end{figure}

\end{document}